\documentclass[%
aps,%
prl,%
preprint,%
showpacs,
groupedaddress]{revtex4}

\renewcommand{\ol}{\overline}

\newcommand{\eqn}[1]{&\hspace{-0.8em}#1\hspace{-0.8em}&}

\newcommand{\abs}[1]{\mbox{$\left| #1 \right|$}}
\newcommand{\eV}{\mbox{eV}}
\newcommand{\keV}{\mbox{keV}}
\newcommand{\GeV}{\mbox{GeV}}

\begin{document}

\title{The $\nu$MSM, Dark Matter and Neutrino Masses}

\author{Takehiko Asaka, Steve Blanchet, and Mikhail Shaposhnikov}

\affiliation{%
Institut de Th\'eorie des Ph\'enom\`enes Physiques,
Ecole Polytechnique F\'ed\'erale de Lausanne,
CH-1015 Lausanne, Switzerland}

\date{March 7, 2005}

\begin{abstract}

We investigate an extension of the Minimal Standard Model by
right-handed neutrinos (the $\nu$MSM) to incorporate neutrino
masses consistent with oscillation experiments.  Within this
theory, the only candidates for dark matter particles are sterile
right-handed neutrinos with masses of a few keV. Requiring that
these neutrinos explain entirely the (warm) dark matter, we find
that their number is at least three. We show that, in the minimal
choice of three sterile neutrinos, the mass of the lightest active
neutrino is smaller than ${\cal O}(10^{-5})$~eV, which excludes
the degenerate mass spectra of three active neutrinos and fixes
the absolute mass scale of the other two active neutrinos.

\end{abstract}


\maketitle

{\em Introduction.}---
In the past decade, neutrino experiments have provided convincing
evidence for neutrino masses and mixings.  The anomaly in
atmospheric neutrinos is now understood by $\nu_\mu \rightarrow
\nu_\tau$ oscillation~\cite{Fukuda:1998mi}, while the solar
neutrino puzzle is solved by the oscillation $\nu_e \rightarrow
\nu_{\mu, \tau}$~\cite{Ahmed:2003kj,Eguchi:2003gg} incorporating
the MSW LMA solution \cite{MSW}. Current data are consistent with
flavor oscillations between three active neutrinos~\cite{LSND},
and show that the mass squared differences are $\Delta m_{\rm
atm}^2 = [2.2^{+0.6}_{-0.4}] \cdot 10^{-3}$~eV$^2$ and $\Delta
m_{\rm sol}^2 = [8.2^{+0.3}_{-0.3}] \cdot 10^{-5}$~eV$^2$
\cite{Gonzalez-Garcia:2004jd}. These phenomena demand  physics
beyond the minimal standard model (MSM), and various possibilities
to incorporate neutrino masses in the theory have been proposed
~\cite{NeutrinoMass}. The simplest one is adding ${\cal N}$
right-handed SU(2)$\times$U(1) singlet neutrinos $N_I$
($I=1,\dots,{\cal N}$) with most general gauge-invariant and
renormalizable interactions described by the Lagrangian:
\begin{eqnarray}
  \delta {\cal L}
  = \ol N_I i \partial_\mu \gamma^\mu N_I
  - f^{\nu}_{I\alpha} \, \Phi  \ol N_I L_\alpha
  - \frac{M_I}{2} \; \ol {N_I^c} N_I + h.c. \,,
\end{eqnarray}
where $\Phi$ and $L_\alpha$ ($\alpha=e,\mu,\tau$) are the Higgs
and lepton doublets, respectively, and both Dirac ($M^D = f^\nu
\langle \Phi \rangle$) and Majorana ($M_I$) masses for neutrinos
are introduced. We have taken a basis in which mass matrices of
charged leptons and right-handed neutrinos are real and diagonal.
We shall call this model ``the {\it $\nu$ Minimal Standard Model}
(the $\nu$MSM)'' (not to be confused with ``the new MSM" of
\cite{Davoudiasl:2004be}). This model satisfies all the principles
of quantum field theory which were so successful in the
construction of the MSM. It should be thus thoroughly studied as
the simplest and experimentally-motivated extension of the MSM.

The $\nu$MSM with ${\cal N}$ singlet neutrinos contains quite a
number of free parameters, i.e. Dirac ($M^D_{I,\alpha}$) and
Majorana ($M_I$) masses. For example, for ${\cal N}=2$ the number
of extra real parameters is 11 (2 Majorana masses, 2 Dirac masses,
4 mixing angels and 3 CP-violating phases), whereas for ${\cal
N}=3$ this number is 18 (3 Majorana masses, 3 Dirac masses, 6
mixing angels and 6 CP-violating phases). These parameters can be
constrained by the observation of neutrino oscillations. The
immediate  consequence  of the existence of two distinct scales
$\Delta m_{\rm atm}^2$ and $\Delta m_{\rm sol}^2$ is that the
number of right-handed neutrinos must be ${\cal N} \ge 2$.
However, we know little about the absolute values of masses for
active neutrinos as well as right-handed neutrinos. This is simply
because the oscillation experiments tell us only about the mass
squared differences of active neutrinos.

On the other hand, cosmology can play an important role to restrict
the parameter space of the $\nu$MSM.  Recently, various cosmological
observations have revealed that the universe is almost spatially flat
and mainly composed of dark energy ($\Omega_{\Lambda}=0.73 \pm
0.04$), dark matter ($\Omega_{\rm dm} = 0.22 \pm 0.04$) and baryons
($\Omega_b = 0.044 \pm 0.004$)~\cite{Eidelman:2004wy}. The $\nu$MSM
can potentially explain dark matter $\Omega_{\rm dm}$ and baryon
$\Omega_{b}$ abundances, and can be consistent with the dark energy
requirement via the introduction of a small cosmological constant.

To be more precise, the baryon asymmetry of the universe
($\Omega_b$) can be produced via the leptogenesis
mechanism~\cite{Fukugita:1986hr} or via neutrino oscillations
\cite{Akhmedov:1998qx} with the use of anomalous electroweak
fermion number non-conservation at high
temperatures~\cite{Kuzmin:1985mm}. Furthermore, the $\nu$MSM can
offer a candidate for dark matter.  The present energy density of
active neutrinos is severely constrained from the observations of
the large scale structure.  The recent
analysis~\cite{Seljak:2004xh} shows that the sum of active
neutrino masses should be smaller than 0.42~eV and $\Omega_{\nu}
h^2 \le 4.5 \cdot 10^{-3}$, which is far below the observed
$\Omega_{\rm dm}$.  The unique dark-matter candidate in the
$\nu$MSM is then a right-handed neutrino which is stable within
the age of the universe. Indeed, it has been shown in
~\cite{SterileNeutrinoWDM} - \cite{Abazajian:2001nj} that sterile
right-handed neutrinos with masses of ${\cal O}(1)$ keV are good
candidates for warm dark matter. Note that, in our analysis, we
take the very conservative assumption of the validity of the
standard Big Bang at temperatures below $1$ GeV and disregard the
possibilities of extremely low reheating temperatures of inflation
as  $T_R \lesssim 1$ GeV \cite{Gelmini:2004ah}.

In this letter, we explore the hypothesis that the $\nu$MSM is a
correct low-energy theory which incorporates dark matter. We
demonstrate that the theory with ${\cal N}=2$ fails to do so. We show
that for the choice ${\cal N}=3$ the mass of the lightest active
neutrino $m_1$ is constrained from above by the value ${\cal
O}(10^{-5})$ eV, and therefore, that the masses of other neutrinos
are fixed to be $m_2 = \sqrt{\Delta m_{\rm sol}^2}$ and $m_3 =
\sqrt{\Delta m_{\rm atm}^2 + \Delta m_{\rm sol}^2}$ in the normal
or $m_2 = \sqrt{\Delta m_{\rm atm}^2}$ and $m_3 = \sqrt{\Delta m_{\rm
atm}^2 + \Delta m_{\rm sol}^2}$ in the inverted hierarchy of
neutrino masses, respectively. This rejects the possibility that all
active neutrinos are degenerate in mass. In other words, for a most
natural choice of ${\cal N}=3$, the cosmological observation of dark
matter allows one to make a (potentially) testable prediction on the
active neutrino masses and on the existence of a sterile neutrino
with a mass in the keV range. We stress that these results are valid
in spite of a large number of free parameters of the $\nu$MSM.
Finally, for ${\cal N}\geq 4$, no model-independent extra constraints
on the masses of active neutrino can be derived.

{\em Neutrino Masses and Mixing.}---
Let us first discuss neutrino masses and mixing in the $\nu$MSM.
We will restrict ourselves to the region in which the Majorana
neutrino masses are larger than the Dirac masses, so that the
seesaw mechanism~\cite{Seesaw} can be applied. Note that this does
not reduce generality since the latter situation automatically
appears when we require the sterile neutrinos to play a role of
dark matter, as we shall see. Then, right-handed neutrinos $N_I$
become approximately the mass eigenstates with $M_1 \le M_2 \le
\dots \le M_{\cal N}$, while other eigenstates can be found by
diagonalizing the mass matrix:
\begin{eqnarray}
  \label{eq:Mseesaw}
  M^\nu = \left(M^D\right)^T \; M_I^{-1} \; M^D \,.
\end{eqnarray}
which we call the seesaw matrix.  The mass eigenstates
$\nu_i$ ($i=1,2,3$) with $m_1 \le m_2 \le m_3$ are found from
\begin{eqnarray}
  \label{eq:Mseesawdiag0}
  U^T M^\nu U = M^\nu_{\rm diag} =
  \mbox{diag}(m_1, m_2, m_3 ) \,,
\end{eqnarray}
and the mixing in the charged current is expressed by
$\nu_\alpha = U_{\alpha i} \, \nu_i + \Theta_{\alpha I}\, N_I^c$
where $\Theta_{\alpha I} = (M^D)^\dagger_{\alpha I} M_I^{-1} \ll
1$ under our assumption.  This is the reason why right-handed
neutrinos $N_I$ are often called ``sterile'' while $\nu_i$
``active''.

As explained before, $\Delta m_{\rm atm}^2$ and $\Delta m_{\rm
sol}^2$ require the number of sterile neutrinos ${\cal N} \ge 2$.
For the minimal choice ${\cal N}=2$, one of the active neutrinos
is exactly massless ($m_1=0$).  For ${\cal N} \ge 3$ the smallest
mass can  be in the range $0 \le m_1 \lesssim {\cal O}(0.1)$
eV~\cite{Seljak:2004xh}.   In particular, the degenerate mass
spectra of active neutrinos are possible when $m_1{}^2 \gtrsim
\Delta m_{\rm atm}^2$. Note also that there are two possible
hierarchies in the masses of active neutrinos, i.e. $\Delta m_{\rm
atm}^2 = m_3^2 - m_2^2$ ($m_2^2 -m_1^2$) and $\Delta m_{\rm sol}^2
= m_2^2 - m_1^2$ ($m_3^2 - m_2^2$) in the normal (inverted)
hierarchy.

{\em Sterile Neutrino as Warm Dark Matter.}---
In the $\nu$MSM, the only candidates for dark matter are the
long-lived sterile neutrinos.  Let us discuss here the
requirements for this scenario.

A sterile neutrino, say $N_1$, decays mainly into three active
neutrinos in the interesting mass range $M_1 \ll  m_e$
(see Eq.~(\ref{eq:DMMass}) below) and its lifetime
is estimated as \cite{Dolgov:2000ew}
\begin{eqnarray}
  \tau_{N_1} = 5 \times 10^{26}\,\mbox{sec}
  \left( \frac{M_1}{1~\keV} \right)^{-5}
  \left( \frac{\ol \Theta^2}{10^{-8}} \right)^{-1} \,,
\label{lifetime}
\end{eqnarray}
where we have taken $\abs{\Theta_{\alpha 1}} = \ol \Theta$ for
$\alpha = e,\mu,\tau$. We can see that it is stable within the age
of the universe $\sim 10^{17}$~sec in some region of the parameter
space ($M_1$,$\Theta$).

When the active-sterile neutrino mixing $\abs{\Theta_{\alpha I}}$ is
sufficiently small, the sterile neutrino $N_I$ has never been in
thermal equilibrium and is produced in non-equilibrium reactions.
The production processes include various particle decays and
conversions of active into sterile neutrinos (see Ref.~\cite{ABS}).
The dominant production mechanism is due to the active-sterile
neutrino
oscillations~\cite{Dodelson:1993je,Dolgov:2000ew,Abazajian:2001nj},
and the energy fraction of the present universe from the sterile
neutrino(s) is~\cite{Abazajian:2001nj}
\begin{eqnarray}
  \label{eq:OmegaN1}
  \Omega_{N}h^2 \sim 0.1
  \sum_I \sum_{\alpha=e,\mu,\tau}
  \left( \frac{ |\Theta_{\alpha I}|^2 }{10^{-8}} \right)
  \left( \frac{ M_I }{1~\keV} \right)^2 \,,
\end{eqnarray}
where the summation of $I$ is taken over the sterile neutrino
$N_I$ being dark matter.  The most effective production occurs
when the temperature is \mbox{$T_* \simeq 130\,\mbox{MeV}
(M_I/1~\keV)^{1/3}$~\cite{Barbieri:1989ti,Dodelson:1993je}}.
Here we assumed for simplicity the flavor universality among
leptons in the hot plasma, which is actually broken since $T_\ast
\le m_\tau$.  However, its effect does not alter our final
results. Further, we have taken the lepton asymmetry at the
production time to be small ($\sim 10^{-10}$), which is a most
conservative assumption. In this case there is no resonant
production of sterile neutrinos coming from large lepton
asymmetries~\cite{Shi:1998km,Abazajian:2001nj}. We therefore find
from the definition of $\Theta$ that the correct dark-matter
density is obtained if
\begin{eqnarray}
  \label{eq:DMCondition}
  \sum_I \sum_{\alpha = e,\mu,\tau} \abs{ M^D_{I \alpha}}^2
  = m_0^2 \,,
\end{eqnarray}
where $m_0 = {\cal O}(0.1)$ eV.  Notice that this constraint on
dark-matter sterile neutrinos is independent of their masses, at
least for $M_I$ in the range discussed below.

The sterile neutrino, being warm dark matter, further receives
constraints from various cosmological observations and the
possible mass range is very restricted as
\begin{eqnarray}
  \label{eq:DMMass}
  2~ \keV \lesssim M_I \lesssim 5~ \keV \,,
\end{eqnarray}
where the lower bound comes from the cosmic microwave background
and the matter power spectrum inferred from Lyman-$\alpha$ forest
data~\cite{Viel:2005qj}, while the upper bound is given by the
radiative decays of sterile neutrinos in dark matter halos limited
by X-ray observations~\cite{Abazajian:2001vt}. These constraints are
somewhat stronger than the one coming from Eq. (\ref{lifetime}).

{\em Consequence of Sterile Neutrino Dark Matter.}---
We have found that the hypothesis of sterile neutrinos being warm
dark matter is realized in the $\nu$MSM when the two constraints
(\ref{eq:DMCondition}) and (\ref{eq:DMMass}) are satisfied. We
shall now see that they put important bounds on the number of
sterile neutrinos and on the masses of the active ones. To find
them, let us first rewrite the diagonalized seesaw mass matrix
(\ref{eq:Mseesawdiag0}) in the form
\begin{eqnarray}
  \label{eq:Mseesawdiag}
  M^\nu_{\rm diag} = S_1 + \dots + S_{\cal N} \,,
\end{eqnarray}
where $S_I$ denotes a contribution from each sterile neutrino and is
given by $(S_I)_{ij} = X_{Ii} X_{Ij}$ with $X_{Ii} = (M^D \, U)_{I
  i}/\sqrt{M_I}$.  Note that each matrix satisfies the relation $\det
S_I = \det(S_I+S_J) = 0$ from its construction.  The condition
(\ref{eq:DMCondition}) is then written as
\begin{eqnarray}
  \label{eq:DMConst2}
  \sum_I \sum_{i=1}^3  \frac{M_I}{M_1} \abs{X_{I i}}^2
  = \frac{m_0^2}{M_1} \equiv m_{\nu}^{\rm dm} \,,
\end{eqnarray}
and the mass range in Eq.~(\ref{eq:DMMass}) gives
\begin{eqnarray}
  \label{eq:MnuDM}
  m_\nu^{\rm dm} = {\cal O}(10^{-5}) \eV\,.
\end{eqnarray}

First of all, let us show that the minimal possibility ${\cal
N}=2$ cannot satisfy the dark-matter constraints and the
oscillation data simultaneously. In this case, the lightest active
neutrino becomes massless ($m_1 = 0$). By taking the trace of both
sides in Eq.~(\ref{eq:Mseesawdiag}), we find that
\begin{eqnarray}
  m_2 + m_3 = \sum_{i=1}^3 ( X_{1i}{}^2 + X_{2i}{}^2 ) \,.
\end{eqnarray}
This equation must hold for both real and imaginary parts.
When both sterile neutrinos $N_1$ and $N_2$ are assumed to be dark matter,
the condition (\ref{eq:DMConst2}) together with $M_1$ and $M_2$
in Eq.~(\ref{eq:DMMass}) leads to
\begin{eqnarray}
  m_2 + m_3 \le \sum_{i=1}^3 ( \abs{X_{1i}}^2 + \abs{X_{2i}}^2 )
  \leq m_{\nu}^{\rm dm}\,.
\end{eqnarray}
This inequality cannot be satisfied since $m_\nu^{\rm dm} ={\cal
O}(10^{-5})$~eV and $m_3 = \sqrt{\Delta m_{\rm atm}^2 + \Delta
m_{\rm sol}^2} \simeq 5 \cdot 10^{-2}$~eV from neutrino
oscillations.

Further, when only one of two sterile neutrinos, say $N_1$, is
assumed to be dark matter, its Dirac Yukawa couplings are
restricted as shown in Eq.~(\ref{eq:DMConst2}). Although the
couplings of $N_2$ can be taken freely, they are not important for
our discussion.  What we shall use here is the simple fact that
the determinant of the matrix $S_2$ in Eq.~(\ref{eq:Mseesawdiag})
is zero.  Then, the equation $\det(S_2) = \det ( M^\nu_{\rm diag}
- S_1 ) = 0$ induces $X_{11}{}^2 \, m_2 \, m_3 = 0$, which is
satisfied only  if \mbox{$X_{11}=0$} since $m_{2,3} \neq 0$ from
the oscillation data. This means that the first row and column of
$S_1$ vanish, and  the matrix $S_2$ should have the same structure
($X_{21}=0$)  because $M_{\nu}^{\rm diag}$ is diagonal and
$m_1=0$. Then, Eq.~(\ref{eq:Mseesawdiag}) is reduced to that for
$2 \times 2$ matrices:
\begin{eqnarray}
  \mbox{diag} (m_2,m_3)
  = X_{1i} X_{1j} + X_{2i} X_{2j} ~~~(i,j=2,3) \,.
\end{eqnarray}
The vanishing determinant of the second matrix on the right-hand
side leads to
\begin{eqnarray}
  m_2 = X_{12}{}^2 + \frac{m_2}{m_3} \, X_{13}{}^2 \,.
\end{eqnarray}
By taking into account the dark matter constraint $\abs{X_{12}}^2 +
\abs{X_{13}}^2 = m_\nu^{\rm dm}$, we obtain the upper bound on~$m_2$:
\begin{eqnarray}
  m_2 \le m_{\nu}^{\rm dm} \,.
\end{eqnarray}
This inequality is inconsistent with $m_\nu^{\rm dm}$ in
Eq.~(\ref{eq:MnuDM}) and $m_2 = \sqrt{ \Delta m_{\rm sol}^2 }
\simeq 9 \cdot 10^{-3}$ eV or $\sqrt{ \Delta m_{\rm atm}^2 }$ for
the normal or inverted hierarchy cases, respectively.  The same
discussion can be applied to the case when only the heavier
sterile neutrino $N_2$ is dark matter.  Therefore, we have shown
that in the ${\cal N}=2$ $\nu$MSM
the requirements on dark matter conflict with the oscillation data.

We then turn to discuss the case ${\cal N}=3$. First, when all three
sterile neutrinos play a role of dark matter simultaneously,
the real part of the trace of Eq.~(\ref{eq:Mseesawdiag}) gives
\begin{eqnarray}
  m_1 + m_2 + m_3 \leq \sum_{I=1}^3  \sum_{i=1}^3 |X_{Ii}{}|^2
  \leq m_\nu^{\rm dm} \,,
\end{eqnarray}
where the final inequality comes from the dark matter constraint
(\ref{eq:DMConst2}) as in the previous case. Although we do not
know the overall scale of $m_i$ from the oscillation data, the
heaviest one $m_3$ should be larger than $\sqrt{\Delta m_{\rm
atm}^2}$ in any case.  Then, this inequality cannot be satisfied
by $m_\nu^{\rm dm}$ in Eq.~(\ref{eq:MnuDM}) and this situation is
excluded.

Next, we consider the case when two of the three sterile
neutrinos, say $N_1$ and $N_2$, are dark matter. In this case,
from the real part of the trace of Eq.~(\ref{eq:Mseesawdiag}), we
find that
\begin{eqnarray}
  m_1 + m_2 + m_3 \leq m_\nu^{\rm dm} + \sum_{i=1}^3 \mbox{Re} X_{3i}{}^2 \,,
\end{eqnarray}
and thus $\sum \mbox{Re} X_{3i}{}^2 > m_3$  since $m_\nu^{\rm dm} \ll
\sqrt{\Delta m_{\rm sol}^2} \le m_2$.  On the other hand,
it is found from $\det ( S_1 + S_2 ) = \det ( M^\nu_{\rm diag} - S_3) = 0$
that, if $m_1 \neq 0$,
\begin{eqnarray}
  \label{eq:Det32}
  1= \frac{X_{31}{}^2}{m_1} + \frac{X_{32}{}^2}{m_2} + \frac{ X_{33}{}^2}{m_3} \,.
\end{eqnarray}
However, this equation cannot be satisfied, since the real part of
the right-hand side is bounded from below as
\begin{eqnarray}
  \frac{\mbox{Re} X_{31}{}^2}{m_1}
  + \frac{\mbox{Re} X_{32}{}^2}{m_2}
  + \frac{\mbox{Re} X_{33}{}^2}{m_3}
  >
  \frac{\sum \mbox{Re} X_{3i}{}^2}{m_3}
  > 1 \,.
\end{eqnarray}
If $m_1 = 0$, $\det( M^\nu_{\rm diag} - S_3)=0$ gives us
$X_{31}=0$.
This results in that $M^\nu_{\rm diag}$ and $S_3$ as well as $(S_1+S_2)$
are reduced to $2 \times 2$ matrices, which verify
$\det S_3 = \det( M^\nu_{\rm diag} - S_1 - S_2)=0$,~i.e.
\begin{eqnarray}
  \eqn{}\left( m_2 - X_{12}{}^2 - X_{22}{}^2 \right)
  \left( m_3 - X_{13}{}^2 - X_{23}{}^2 \right)
  \nonumber \\
  \eqn{} ~~~~ = (X_{12}X_{13}+X_{22}X_{23})^2 \,.
\end{eqnarray}
This equation cannot be satisfied by $X_{Ii}$ restricted by the
dark matter constraint (\ref{eq:DMConst2}). Thus, this case is
also excluded in either $m_1=0$ or $m_1\neq0$ situations.

Finally, let us consider the remaining possibility, i.e. assume
that only one sterile neutrino (e.g. $N_1$) becomes a dark matter
particle. In this case, we also note that $\det( S_2 + S_3 ) =
\det(M^{\nu}_{\rm diag} - S_1 )= 0$, which induces
\begin{eqnarray}
  m_1 = X_{11}{}^2 + \frac{m_1}{m_2} \, X_{12}{}^2
  + \frac{m_1}{m_3} \, X_{13}{}^3 \,.
\end{eqnarray}
Now, the dark matter constraint (\ref{eq:DMConst2}) takes the
form: $\sum_{i=1}^3 \abs{X_{1i}}^2 = m_\nu^{\rm dm}$. It is then
found that the lightest active neutrino should verify
\begin{eqnarray}
  m_1 \le m_{\nu}^{\rm dm} \,.
\end{eqnarray}
This shows that, when ${\cal N}=3$, there exists a region in the
parameter space of the $\nu$MSM consistent with the observation of
neutrino oscillations and in which one of sterile neutrinos
becomes the warm dark matter of the universe. Finally, we should
stress here that the above argument holds independently of the
mixing angles of neutrinos in $U$.

If the number of sterile neutrinos is greater than the number of
fermionic generations, no general constraints on the masses of
active neutrinos can be derived, since extra sterile neutrinos may
be almost decoupled from the active neutrinos and thus do not
contribute to the seesaw formula. At the same time, they can
easily satisfy the dark matter constraint.

{\em Conclusions.}---
Let us summarize the obtained results.  First, we have shown that
the $\nu$MSM can explain the dark matter in the universe only
provided ${\cal N} \ge 3$, although the neutrino oscillation
experiments allow ${\cal N}=2$.  Interestingly, in this successful
and minimal scenario with ${\cal N} = 3$, the number of sterile
neutrinos is the same as the number of families of quarks and
leptons. Second, in the ${\cal N} = 3$ case, the mass of the
lightest active neutrino should lie in the range $m_1 \le
m_{\nu}^{\rm dm} ={\cal O}(10^{-5})$~eV, which is much smaller
than $\sqrt{\Delta m_{\rm sol}^2}$.  This clearly excludes the
possibility that three active neutrinos are degenerate in mass and
fixes their masses to be $m_3 = [4.8^{+0.6}_{-0.5}] \cdot 10^{-2}$
eV and $m_2= [9.05^{+0.2}_{-0.1}]\cdot 10^{-3}$eV
($[4.7^{+0.6}_{-0.5}]\cdot 10^{-2}$eV) in the normal (inverted)
hierarchy.  An experimental test of the ${\cal N} = 3$ $\nu$MSM
origin of dark matter would be the discovery of a keV sterile
neutrino by the X-ray observatories~\cite{Abazajian:2001vt} and
the finding of the active neutrino masses in the predicted range.

Finally, we should mention that the sterile neutrinos irrelevant
to dark matter can be responsible for the baryon asymmetry of the
universe through leptogenesis ~\cite{Fukugita:1986hr} or neutrino
oscillations \cite{Akhmedov:1998qx}. These considerations would
restrict further the parameter space of the $\nu$MSM.  For
example, the conventional thermal
scenario~\cite{Buchmuller:2005eh} works when the lightest among
them is about $10^{10}$ GeV.  The other scenario using neutrino
oscillations requires masses of $100~\GeV \gg M_I \ge 1$
GeV~\cite{Akhmedov:1998qx}.

{\em Acknowledgments.}---
This work has been supported by the Swiss Science Foundation and by
the Tomalla foundation.



\begin{thebibliography}{99}
%
\bibitem{Fukuda:1998mi}
Y.~Fukuda {\it et al.}  [Super-Kamiokande Collaboration],
Phys.\ Rev.\ Lett.\  {\bf 81}, 1562 (1998)
[arXiv:hep-ex/9807003].
%
\bibitem{Ahmed:2003kj}
S.~N.~Ahmed {\it et al.}  [SNO Collaboration],
Phys.\ Rev.\ Lett.\  {\bf 92}, 181301 (2004)
[arXiv:nucl-ex/0309004].
%
\bibitem{Eguchi:2003gg}
K.~Eguchi {\it et al.}  [KamLAND Collaboration],
Phys.\ Rev.\ Lett.\  {\bf 92}, 071301 (2004)
[arXiv:hep-ex/0310047].

\bibitem{MSW} L. Wolfenstein, Phys. Rev. D{\bf 17}, 2369 (1978);
 Phys. Rev. D{\bf 20}, 2634 (1979);
S. P. Mikheyev and A. Yu. Smirnov, Sov. J. Nucl. Phys.
{\bf 42}, 913 (1985),  Nuovo Cim. {\bf C9}, 17 (1986);
 Sov. Phys. JETP {\bf 64}, 4  (1986)

\bibitem{LSND}
We do not include here the LSND anormaly~\cite{Aguilar:2001ty},
which will be tested in the near future~\cite{MiniBooNE}.
%
\bibitem{Aguilar:2001ty}
A.~Aguilar {\it et al.}  [LSND Collaboration],
Phys.\ Rev.\ D {\bf 64}, 112007 (2001)
[arXiv:hep-ex/0104049].
%
\bibitem{MiniBooNE}
[BooNE collaboration], http://www-boone.fnal.gov/
%
\bibitem{Gonzalez-Garcia:2004jd}
See, for an recent analysis,
M.~C.~Gonzalez-Garcia,
arXiv:hep-ph/0410030.

\bibitem{NeutrinoMass}
For example, see reviews,
S.~F.~King,
Rept.\ Prog.\ Phys.\  {\bf 67}, 107 (2004)
[arXiv:hep-ph/0310204];
G.~Altarelli and F.~Feruglio,
New J.\ Phys.\  {\bf 6}, 106 (2004)
[arXiv:hep-ph/0405048].

\bibitem{Davoudiasl:2004be}
H.~Davoudiasl, R.~Kitano, T.~Li and H.~Murayama,
arXiv:hep-ph/0405097.

\bibitem{Eidelman:2004wy}
S.~Eidelman {\it et al.}  [Particle Data Group Collaboration],
Phys.\ Lett.\ B {\bf 592} (2004) 1.

\bibitem{Fukugita:1986hr}
M.~Fukugita and T.~Yanagida,
Phys.\ Lett.\ B {\bf 174}, 45 (1986).

\bibitem{Akhmedov:1998qx}
E.~K.~Akhmedov, V.~A.~Rubakov and A.~Y.~Smirnov,
Phys.\ Rev.\ Lett.\  {\bf 81}, 1359 (1998)
[arXiv:hep-ph/9803255].

\bibitem{Kuzmin:1985mm}
V.~A.~Kuzmin, V.~A.~Rubakov and M.~E.~Shaposhnikov,
Phys.\ Lett.\ B {\bf 155}, 36 (1985).

\bibitem{Seljak:2004xh}
U.~Seljak {\it et al.},
arXiv:astro-ph/0407372.

\bibitem{SterileNeutrinoWDM}
P.~J.~E.~Peebles,
Astrophys.\ J.\  {\bf 258}, 415 (1982);
%
K.~A.~Olive and M.~S.~Turner,
Phys.\ Rev.\ D {\bf 25}, 213 (1982).
%
\bibitem{Dodelson:1993je}
S.~Dodelson and L.~M.~Widrow,
Phys.\ Rev.\ Lett.\  {\bf 72}, 17 (1994)
[arXiv:hep-ph/9303287].
%
\bibitem{Shi:1998km}
X.~d.~Shi and G.~M.~Fuller,
Phys.\ Rev.\ Lett.\  {\bf 82}, 2832 (1999)
[arXiv:astro-ph/9810076].
%
\bibitem{Dolgov:2000ew}
A.~D.~Dolgov and S.~H.~Hansen,
Astropart.\ Phys.\  {\bf 16}, 339 (2002)
[arXiv:hep-ph/0009083].
%
\bibitem{Abazajian:2001nj}
K.~Abazajian, G.~M.~Fuller and M.~Patel,
Phys.\ Rev.\ D {\bf 64}, 023501 (2001)
[arXiv:astro-ph/0101524].
%
\bibitem{Gelmini:2004ah}
G.~Gelmini, S.~Palomares-Ruiz and S.~Pascoli,
Phys.\ Rev.\ Lett.\  {\bf 93}, 081302 (2004)
[arXiv:astro-ph/0403323].

\bibitem{Seesaw}
T.~Yanagida, in {\em Proc. of the Workshop on ``The Unified Theory
and the Baryon Number in the Universe''}, Tsukuba, Japan, Feb.~13-14, 1979,
p.~95, eds. O.~Sawada and S.~Sugamoto, (KEK Report KEK-79-18, 1979, Tsukuba); %
Progr.\ Theor.\ Phys.\ {\bf 64}, 1103 (1980); %
P.~Ramond, in {\em Talk given at the Sanibel Symposium''}, Palm Coast, Fla.,
Feb.~25-Mar.~2, preprint CALT-68-709.
%
\bibitem{ABS}
T.~Asaka, S.~Blanchet and M.~Shaposhnikov,
in preparation.

\bibitem{Barbieri:1989ti}
R.~Barbieri and A.~Dolgov,
Phys.\ Lett.\ B {\bf 237}, 440 (1990);
K.~Kainulainen,
{\em ibid.} {\bf 244}, 191 (1990).

\bibitem{Viel:2005qj}
M.~Viel, J.~Lesgourgues, M.~G.~Haehnelt, S.~Matarrese and A.~Riotto,
arXiv:astro-ph/0501562.

\bibitem{Abazajian:2001vt}
K.~Abazajian, G.~M.~Fuller and W.~H.~Tucker,
Astrophys.\ J.\  {\bf 562}, 593 (2001)
[arXiv:astro-ph/0106002].

\bibitem{Buchmuller:2005eh}
See e.g. a recent review,
W.~Buchmuller, R.~D.~Peccei and T.~Yanagida,
arXiv:hep-ph/0502169.

\end{thebibliography}
\end{document}